\documentclass[pre,aps,twocolumn,showpacs,superscriptaddress]{revtex4-1} 
\usepackage{amsmath,amssymb}
\usepackage{graphicx}
\usepackage{txfonts}
\usepackage{color}
\usepackage{hyperref}
\usepackage{bbold}
\usepackage{framed} 
\usepackage{wrapfig}
\usepackage[a4paper]{geometry}
\global\long\def\av#1{\left\langle #1 \right\rangle }

\begin{document}

\title{Fluctuating local field method probed for a description of small classical correlated lattices}

\author{Alexey N. Rubtsov}
\affiliation{Russian Quantum Center, Novaya 100, 143025 Skolkovo, Moscow Region, Russia}
\affiliation{Department of Physics, M.V.Lomonosov Moscow State University, Leninskie gory 1, 119991 Moscow, Russia} 
\email{ar@rqc.ru}
\begin{abstract}
Thermal-equilibrated  finite classical lattices are considered as a minimal model of the systems showing  an interplay between low energy collective fluctuations and single site degrees of freedom. Standard local field approach, as well as classical limit of the bosonic DMFT method, do not provide a satisfactory description of Ising and Heisenberg small lattices subjected to an external polarizing field. We show that a dramatic improvement can be achieved within a simple approach, in which the local field appears to be a fluctuating quantity related to the low energy degree(s) of freedom.
\end{abstract}
\maketitle

\section {Introduction} 

Lattice ensembles subjected to thermal fluctuations appear widely in different branches of physics, as they can be seen as prototypical models
describing phase transitions in correlated media.  In some special cases the exact solutions are known \cite{Baxter}, while Monte Carlo simulations \cite{Binder} deliver accurate data for generic classical ensembles.
Methodologically, lattice models are exploited as playground for (semi)analytical approaches,
aiming to determine their phase diagram and/or critical properties.  

Statistical properties of a fluctuating ensemble
are typically characterized by two different characteristic length scales \cite{JinnJustin}. Universal long-range scaling behavior takes the place in a vicinity of the transition point. 
Model specific short-range fluctuations occur at typical length of a few lattice periods with a temperature/parameter range remarkably broader than the critical region. A simple way to address short-range fluctuations separately is to consider a finite lattice instead of the thermodynamical limit. For small enough systems, the low energy lattice excitations show up as fluctuations of just a single mode corresponding to the order parameter of the lattice. However the dispersion of this mode is determined by the high energy fluctuations. So the main ingredient of the strong correlations -- mutual influence of different energy scales -- is present. This makes studies of the  statistical properties of finite lattices conceptually important.

\begin{figure}[t]
 \center{\includegraphics[width=1\linewidth]{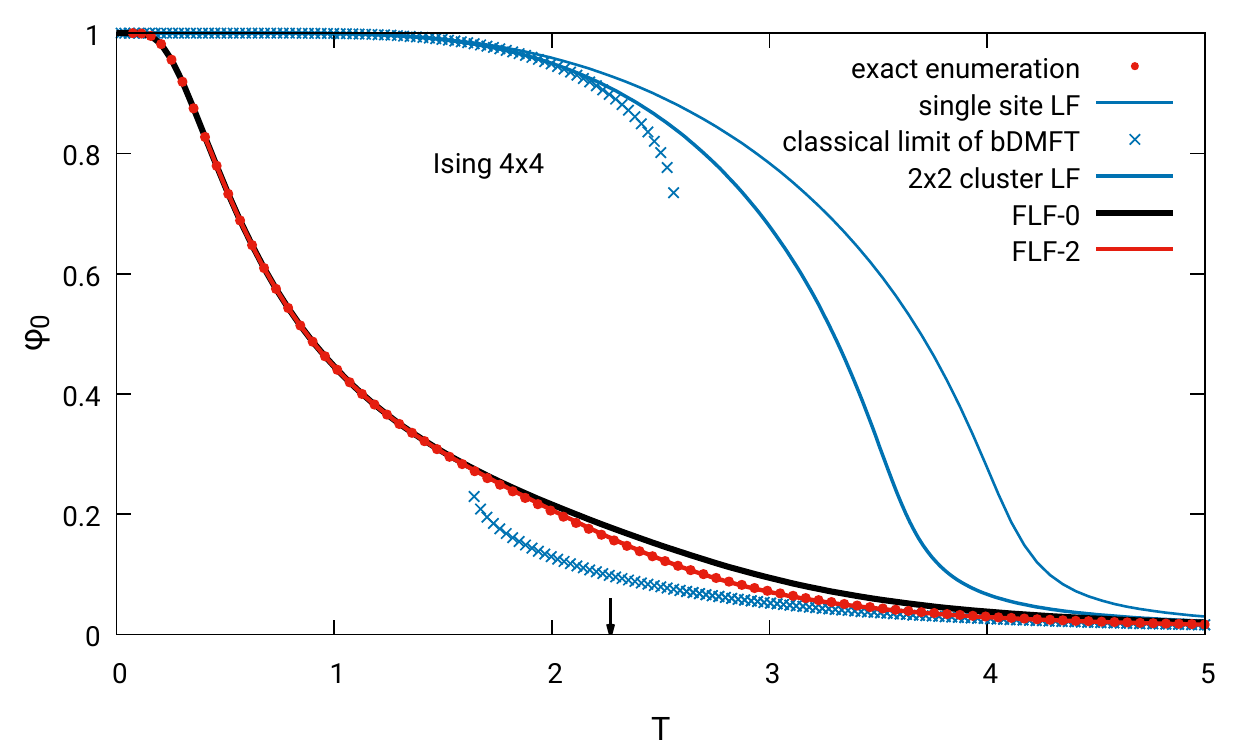}} 
  \center{\includegraphics[width=1\linewidth]{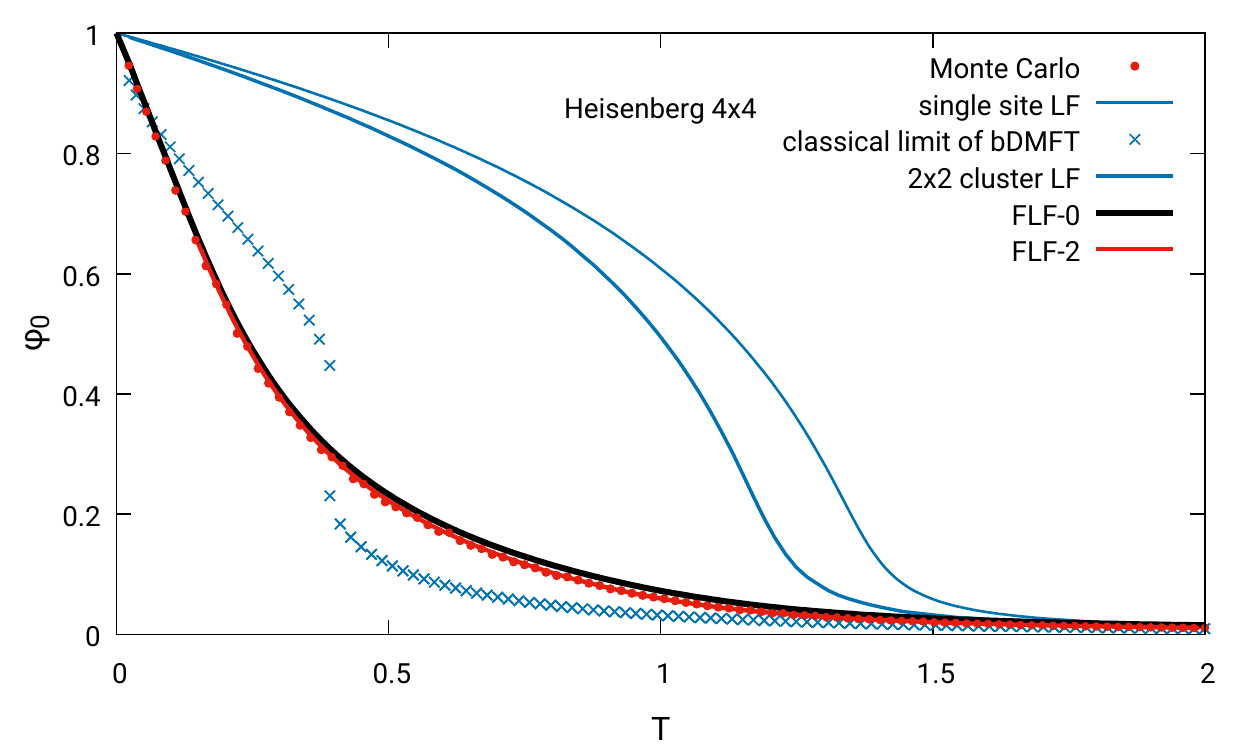}} 
\centering{}\caption{Thermal dependence of the polarization induced in $4\times4$ periodized Ising and Heisenberg lattices by the external field $h=0.03$. Single site Local Field approximation, its cluster generalization, classical version of the bosonic DMFT scheme,  and Fluctuating Local Field method results are compared with the numerically exact reference data. For the Ising system, the value of critical temperature in the thermodynamical limit is marked with arrow.}
\label{figvam} 
\end{figure}

In this paper we ask whether a polarization of a small classical lattice by a finite external field can be 
described within a simple approximation. As an example, Ising and Heisenberg lattices are considered. Our findings are summarized in Figure \ref{figvam}, that presents the numerically exact data and several approximate temperature dependencies of the order parameter. The curves show that 
commonplace mean field methods are not appropriate here, as they give very inaccurate results. We present a Fluctuating Local Field (FLF) method which dramatically improves the situation -- it allows for a quantitatively good description of the lattice polarization in a wide temperature range. The curves labeled FLF-0 and FLF-2 refer to the FLF approximations of a different order. In both cases the FLF is ``almost analytical'', that is  it requires a numerical estimation of just  a single integral.

 Being applied to Ising and Heisenberg lattices, the FLF method is of a methodological importance only, because straightforward Monte Carlo simulations are technically easy for these models. We however expect that the approach can be helpful for more sophisticated cases, including modeling of quantum ensembles, where Monte Carlo simulations are complicated. Indeed,  a solution of so-called cluster problem is a key ingredient of many theories describing an infinite lattice.   These approaches to lattice problems require a calculation of the cluster response to an effective field defined in a self-consistent way. In this paper, we provide an example of such a consideration. Also, solution of the cluster problem can be used to integrate out the high-energy degrees of freedom preceding the renormalization-group procedure.
For example a seminal Kadanov's block renormalization \cite{Kadanoff} uses supercells on the lattice, each containing a number of original lattice cells.
Then integrating out the lattice degrees of freedom within a supercell means calculating its Landau free energy $F(\phi_0)$, where $\phi_0$ is a local order parameter: one sums over possible configurations of lattice fields belonging to the supercell preserving the single degree of freedom $\phi_0$.  
An input data for the functional renormalization group (FRG) treatment \cite{FRG},  one can use  a Legendre transform of $F(\phi_0)$ or, equally,  cluster response to the applied  field.  A problem of the  response of finite lattices to an applied field is also interesting in its own, for example in the context of the molecular superparamagnetism \cite{superpara}.

\section{Model}

We consider periodized two-dimensional Ising and Heisenberg lattices placed in a small but finite external field $h$. For the $N$-site lattice equilibrated at temperature $T$, the partition function  reads
\begin{equation}\label{Z}
Z=\int \exp{ \left( \frac{1}{T}\sum_{<jj'>} (s_j s_{j'}) +\frac{1}{T} \sum_j (h s_j) \right)} d^N s.
\end{equation}
The constraint $|s_j|^2=1$ is imposed for scalar (Ising) or 3-component vector (Heisenberg) fields, so that for the Ising model (\ref{Z}) contains just a summation over $s_j=\pm 1$.

Figure \ref{figvam} shows the  temperature dependence of the order parameter $\phi_0=N^{-1}\sum_j \av{s_j}$ for the $4\times 4$  periodized lattices placed  in the external field $h=0.03$. Reference data for the Ising and Heisenberg system are respectively obtained from the exact enumeration over $2^{16}$ possible states and using a Monte Carlo procedure.
As one can see, the response inherits the behavior of the corresponding infinite lattice models. The planar Ising model shows a phase transition at the critical temperature $T_c\approx 2.27 $, indicated by the arrow in Figure \ref{figvam}. In a finite system, it corresponds
to a superparamagnetic (i.e.very soft) collective mode formed near the critical temperature.  A value of the applied field was chosen in such a way that 
it induces almost saturated polarization of this mode at low temperature whereas for high temperature the polarization is small. The overall temperature dependence of the order parameter can be seen as a heavily deformed one of an infinite lattice.
For the Heisenberg model in two dimensions, the phase transition at finite temperature is destroyed by fluctuations. Consequently, a formation of the collective mode is observed close to the zero temperature, and the graph for the temperature dependence of the order parameter has a shape different from the Ising case. 

\section{Known local field methods}

The simplest local field (LF) approximation reduces the problem to a single site placed in the effective field:
\begin{equation}\label{Zsite}
z_\nu=\int \exp\left( \frac {(\nu s)}{T}\right) ds.
\end{equation}
The local field self-consistency condition for the order parameter reads 
\begin{equation}\label{MF}
\phi_0= \phi^{site}_{h+4 \phi_0},
\end{equation}
where $\phi^{site}_\nu=T \frac{\partial}{\partial \nu} \ln z_\nu$ is the response of a single lattice site. Physically the LF approximation assumes that correlations are fully localized in space. It is known that this assumption overestimates correlations. The LF results are presented in Figure 1 with thin lines. They show a poor performance of the LF approximation, except for the high temperature region where 
different sites are uncorrelated.


The LF approximation can be extended by taking into account the second moment of the single site problem (\ref{Zsite}). Physically this corresponds to an assumption that the correlations between different sites obey approximately Gaussian statistics. Such an approach can be seeing as a classical limit of so-called bosonic dynamical mean field theory (bDMFT) \cite{bDMFT}, which operates with the local average and Green's function of a bosonic lattice problem. Being formulated for the classical lattice problem (\ref{Z}), the bDMFT operates with a hybridized single site problem
\begin{equation}\label{Zsitehyb}
z_{\nu, \Delta}=\int \exp\left( \frac {(\nu s) -\frac{1}{2}(s \Delta s)}{T}\right) ds,
\end{equation}
where $\Delta$ is a scalar number for the Ising lattice and a 2-rank tensor for the Heisenberg one.
The self consistent bDMFT conditions for $\phi_0, \Delta$ read
\begin{equation}\label{bDMFT}
\begin{array}{l}
\phi_0= \phi^{site}_{\nu, \Delta},\\ \\
g_{\nu, \Delta}=\frac{1}{N}\sum_k \left(g_{\nu, \Delta}^{-1}+\Delta+2(\cos k_x + \cos k_y)\right)^{-1},\\ \\
\nu=h+(4-\Delta) \phi_0,
\end{array}
\end{equation}
where $\phi^{site}_{\nu, \Delta}=T \frac{\partial}{\partial \nu} \ln z_{\nu, \Delta}$ and $g=-  \frac{\partial}{\partial \nu}\phi^{site}_{\nu, \Delta}$ are the first and the second momenta for the hybridized site problem (\ref{Zsitehyb}), and the sum runs over all $k$-vectors of the lattice. 
 
An account of the second local moment allows to extend the high temperature domain where the approximation performs well, as Figure \ref{figvam} shows. However the low temperature region is still described quite poorly. Moreover for the Ising system there is now a coexistence region giving rise to a first order phase transition, which is completely irrelevant for the lattices under study.

Other possible extension of the single site LF method is cluster schemes. Whose methods partly take the nonlocal correlations into account by considering a supercell instead of the single site. There exists a number of 
schemes different in self consistency conditions \cite{cluster}. In the simplest direct-space cluster method with the $2\times 2$ supercell, the order parameter fulfills the equation 
\begin{equation}\label{supercell}
\phi_0= T \frac{\partial }{\partial \nu} \ln \int  \exp{\left(\frac{1}{T}\sum_{<jj'>} (s_j s_{j'}) + \frac{1}{T} \sum_j ( (h + 2  \phi_0) s_j)\right) } d^4s .
\end{equation}
Here the indexes run over the four sites of $2\times 2$ supercell; each of these sites is connected with two outer lattice cites, hence the effective field $2 \phi_0$. The order parameter temperature dependence obtained  from the equation (\ref{supercell}) is shown in Figure \ref{figvam} in dashed lines.
Compared to the single site LF approximation, the scheme again improves the results only for the high temperature region, where the short range fluctuations dominate. The overall advantage over the single site method is not significant, especially taking into account an increased complexity of the approximation.

The examples considered convinced us that the LF family of approximations is not applicable for 
the systems under consideration. This is because the $k=0$ mode exhibits strong non-Gaussian fluctuations in the low temperature regime, and this is not captured by the schemes assuming Gaussian statistics or uncorrelated motion at long range. Therefore we expect that other cluster schemes not considered here 
would not improve the situation drastically. 

We could not find other known  simple and physically transparent approximations being relevant to the problem.
Besides the cluster schemes, long-range spatial nonlocality can be taken into account using the diagrammatic extensions around the LF result  with respect to high-order vortexes of the single site problem \cite{vertex}. Technical complexity of such schemes definitely exceeds the level we would like stay at. For problems with a weak nonlinearity,
say the discrete $\phi^4$ model, one could consider low-order diagrammatic approximations starting from the independent modes in the $k$-space. However, the constrain $s_j^2=1$ formally corresponds to an infinitely large nonlinearity, and this family of approximations  is not applicable.

\section{Fluctuating local field approach}

The main result of this paper is a formulation of the Fluctuation Local Field method and 
its application to classical Ising and Heisenberg finite-size lattices. 
An advantage of the FLF method is that it explicitly considers the $k=0$ mode as a special case,
thus allowing an accurate treatment of the long range correlations.
Simultaneously its  technical complexity is not remarkably increased compared to the LF approximation.

To introduce the FLF method, we express  for the partition function in the following way:
\begin{widetext}
\begin{equation}\label{Zf}
Z=\iint \exp{ \left( \frac{1}{T}\sum_{<jj'>} (s_j s_{j'}) +\frac{h}{T} \sum_j s_j - \frac{N \lambda}{2 T}\left|\sum_j \frac{s_j}{N} - \frac{\nu-h}{\lambda}\right|^2 \right)} d^N s d\nu,
\end{equation}
where the integration variable $\nu$ is a scalar (3-component vector) for Ising (Heisenberg) model, and $\lambda$ is a parameter to be defined later. As it is immediately seen from the integration over $\nu$ in (\ref{Zf}) , the partition function (\ref{Z}) remained unchanged, up to a prefactor. 

We
introduce a Landau free energy for the field $\nu$
\begin{equation}\label{Ff}
F_{\nu,h}= \frac{N  |\nu-h|^2}{2 \lambda} - T \ln \int \exp{ \left( \frac{1}{T}\sum_{<jj'>} (s_j s_{j'}) +\frac{1}{T} \sum_j (\nu s_j) - \frac {\lambda}{2 NT}\left|\sum_j s_j\right|^2   \right)} d^N s,
\end{equation}
\end{widetext}
so that 
\begin{equation}\label{Znu}
Z=\int \exp\left(-\frac{F_{\nu,h}}{T}\right) d\nu.
\end{equation}

We will use the local field estimation of $F_{\nu, h}$ with subsequent integration over $\nu$. This course of action resembles the FRG, but there are two important differences. First, the exponent of (\ref{Ff}) contains the term $\frac{\lambda}{2 N}\left| \sum_j s_j \right|^2$ corresponding to an effective long-range interaction which hardens the $k=0$ mode. In FRG, this term does not appear. Second, we will consider not Taylor coefficients for $F_{\nu}$ near zero, but the entire graph for the whole relevant range of $\nu$.
  
A freedom  of choosing the parameter $\lambda$ can be exploited to obtain a particularly simple scheme.
Indeed, for 
\begin{equation}\label{lambda}
\lambda=4 \left(1-\frac{1}{N}\right)^{-1} 
\end{equation}
the interaction with nearest neighbors in (\ref{Ff}) is in average compensated by the effective long-range interaction, so that  on the local field level the system becomes a set of independent sites.  In this case 
one obtains $F_{h, \nu}\approx F_{h, \nu}^{(0)}$ with
\begin{equation}\label{Ff0}
F_{h, \nu}^{(0)}=N\left( \frac{  |\nu-h|^2}{2 \lambda} - T \ln z_{\nu}\right),
\end{equation}
where $z_\nu$ is defined by the equation (\ref{Zsite}). For the specific models we consider, $z_\nu={\rm cosh} \frac{\nu}{T}$ (Ising) and $z_\nu=\frac{T}{\nu} {\rm sinh} \frac{\nu}{T}$ (Heisenberg).

The approximation (\ref{Ff0}) can be improved using the thermodynamical perturbation theory \cite{LL5}.
Taking the choice (\ref{lambda}) into account, the terms neglected while passing from (\ref{Ff}) to (\ref{Ff0})  can be expressed as   
\begin{equation}\label{W}
W=- \sum_{j\neq j'} \tilde J_{j-j'} (s_j-\phi^{site}) (s_{j'}-\phi^{site}),
\end{equation}
where $\tilde J_{j-j'}$ equals $1-\lambda/N$ for the nearest neighbors and $-\lambda/N$ otherwise. The first-order correction to $F^{(0)}$ with respect to $W$ is vanished. An estimation of the  second-order correction   gives 
\begin{equation}\label{Ff2}
F_{h, \nu}^{(2)}=N\left( \frac{  |\nu-h|^2}{2 \lambda} - T \ln z_{\nu}\right) - \frac{N ||g_\nu||^2}{4 T} \sum_{j} \tilde J_{j}^2,
\end{equation}
where the second-order momentum of the impurity problem $g_\nu$ appeared previously in the equations (\ref{bDMFT}). In the case under consideration the hybridization $\Delta$ is absent. For the Heisenberg model, the quantity $g_\nu$ has a tensor nature because of the anisotropy induced by the field $\nu$. The norm $||g_\nu||^2$ entering (\ref{Ff2}) is a sum of the tensors components' squares.

\begin{figure}[t]
 \center{\includegraphics[width=1\linewidth]{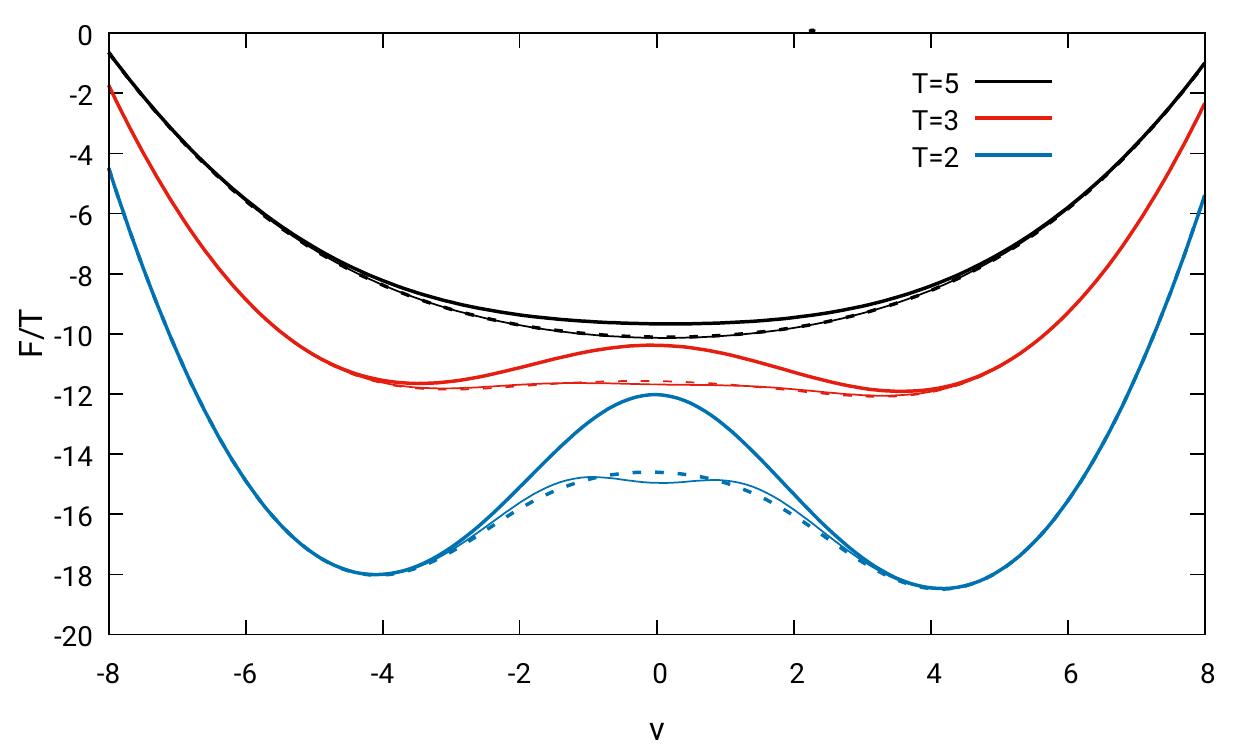}} 
\centering{}\caption{Exact (dashed lined) dependencies for the Landau free energy for the field $\nu$, compared with the FLF-0 (thick lines) and FLF-2 (thin lines) approximations. The data are presented for the $4\times 4$ Ising system at different values of temperature. The curves are slightly tilted because of the  polarizing field $h=0.03$. }
\label{figsdva} 
\end{figure}

Figure \ref{figsdva} shows a comparison of the approximations (\ref{Ff0}) and (\ref{Ff2}) for the $4\times 4$ Ising lattice with the reference data obtained by the exact enumeration in (\ref{Ff}). For several values of temperature, we plot the graphs of $F_{\nu,h}/T$.  Through the paper, we refer to the approximations (\ref{Ff0}) and (\ref{Ff2}) as FLF-0 and FLF-2, respectively.
It can be observed that the FLF-0  performs rather good.
First we note that for any temperature, the approximation matches the exact result for large $\nu$. This indicates that a strong applied field  suppresses fluctuations, so that the theory is exact for $\nu\to \infty$. Next, for a relatively high temperature $T=5$, an accuracy of the approximation is good within the entire range of $\nu$. This is because in the high temperature limit the fluctuations are local, and the approximation becomes exact. For a low temperature $T=2$, the curve has two pronounced minima, slightly different due to a finite external field $h=0.03$. One can see that the minima in fact belong to the large-$\nu$ region where the approximation performs well.   Here we recall that it is important to consider the entire curve for $F_\nu$, as its polynomial approximation based on the first Taylor coefficients near $\nu=0$ would give worse results.   Since (\ref{Znu}) is contributed by the vicinity of minima, their accurate handling means a good performance of the developed theory at low temperatures.  
It is of course not surprising that the theory is valid in the limit cases, but it is remarkable that the high- and low-temperature domains of validity extend quite close to the critical temperature.
The result presented in Figure \ref{figsdva} for $T=3$ can be seen as the worst case; still the deviation from the reference data remains moderate. The FLF-2 scheme 
puts the approximate curves very close to the reference data. The deviations can be seen only for the lowest temperature $T=2$. We have observed that these deviations grow while $T$ is decreased, so that the perturbation theory breaks down at very low temperature. Nevertheless,
it can be successfully applied within the most interesting temperature range $T>0.3$.

The temperature dependence of the order parameter was obtained using the expression $\phi_0=\frac{T}{N}\frac{\partial}{\partial h} \ln Z$, and taking the derivative numerically. For the Heisenberg model, the integration over the angular part of $\nu$ in the expression for $Z$ can be carried out analytically. Therefore both Ising and Heisenberg models require  grid integration over a single variable. The results are presented in Figure \ref{figvam} and demonstrate a very good agreement with the reference data.  The FLF-0 show small deviations at $T\approx 3$ for the Ising model and for $T\approx 0.5 ... 1$ for the Heisenberg one. The FLF-2 results almost coincide with the reference data.
The scheme was found to provide a very good accuracy for the lattices having the size up to about $7\times 7$.
For larger lattices, critical fluctuations are important which have not been not taken into account so far.

\begin{figure}[t]
 \center{\includegraphics[width=1\linewidth]{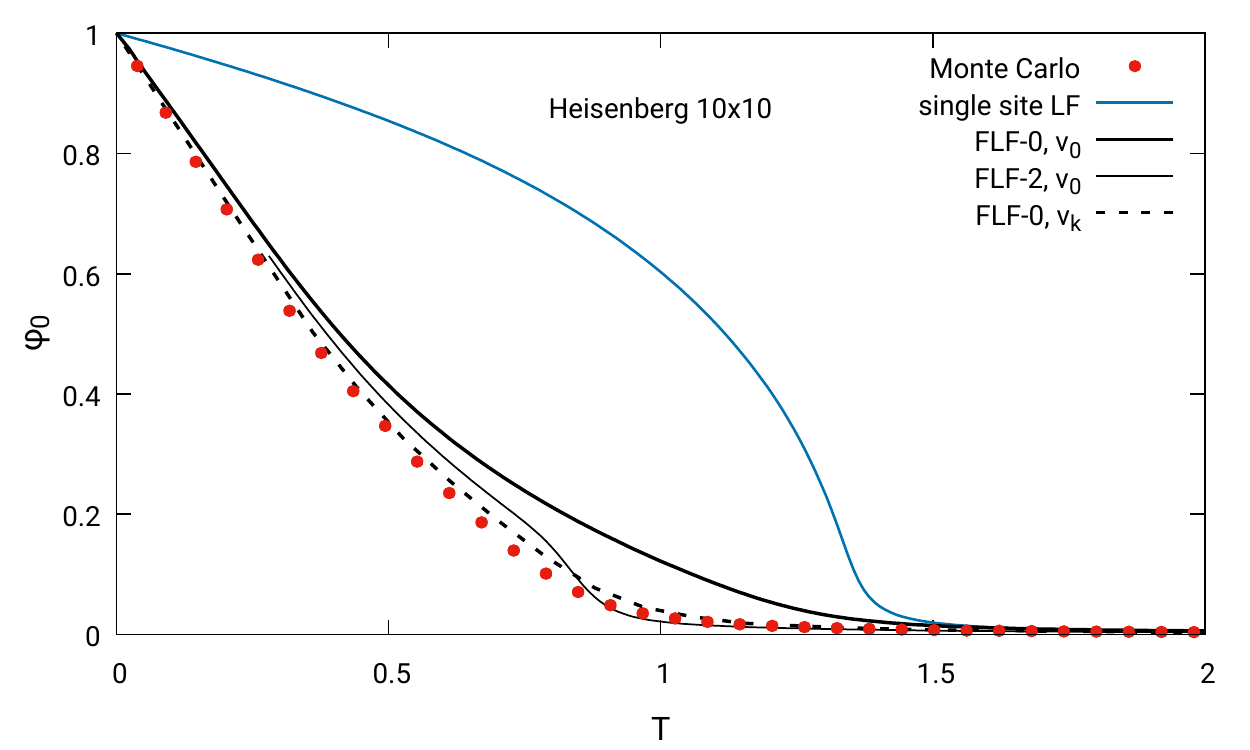}} 
\centering{}\caption{Thermal dependence of the polarization induced in $10\times10$ periodized  Heisenberg lattice by the external field $h=0.01$. Fluctuating Local Field method with the single $k$-mode and for $9$ modes, as well as the Local Field approximation are compared with the numerically exact reference data.}
\label{natroih} 
\end{figure}

Figure \ref{natroih} shows the data for the $10\times 10$ Heisenberg lattice.  Since at small temperatures larger lattices are more sensitive to the applied field
and also to have fluctuation regime more pronounced,  we considered a decreased value of $h=0.01$. Although the FLF method still performs much better than the LF approximation, there is a region where the order parameter is clearly overestimated. The FLF-2 improves the description  from the high-temperature side only. This is because at lower temperature the fluctuations are developed in several low-lying modes of the system, and the uniform fluctuating field $\nu$ does not catch them.  

To take the fluctuations in multiple modes into account, we introduce a $k$-dependent auxiliary field:
\begin{widetext}
\begin{equation}\label{Zf}
Z=\iint \exp{ \left( \frac{1}{T}\sum_{<jj'>} (s_j s_{j'}) +\frac{1}{T} \sum_j (h s_j) - \sum_k \frac{N \lambda_k}{2 T}\left|\sum_j \frac{s_j e^{-i k j}}{N}  - \frac{\nu_k-h \delta_{k,0}}{\lambda_k}\right|^2 \right)} d^N s d\nu,
\end{equation}
\end{widetext}
where $k$ runs over the low energy modes; $\delta_{k,0}$ is a Kronecker delta. For $k\neq 0$  the components $\nu_k$ are complex numbers with the condition $\nu_{k}=\nu_{-k}^*$ imposed.
In an analogy to the previous consideration we take $\lambda_{k}=2 (\cos k_x+ \cos k_y)$. The estimation (\ref{Ff0}) is now replaced with 
\begin{equation}\label{Ffk}
F_{h, \nu}=\sum_k \frac{N  |\nu_k-h \delta_{k,0}|^2}{2 \lambda_k} - T \sum_j \ln z_{\nu_j},
\end{equation}
where $\nu_j=\sum_k \nu_k e^{i k j}$. Formula (\ref{Ffk}) can be called a low-energy effective Hamiltonian, or 
Landau-Ginzburg free energy, for the ``slow'' field $\nu$.
 
For the $10\times 10$ Heisenberg system we have considered the modes with $k_x=0, \pm \frac{2\pi}{10}$
and same for $k_y$, that is nine $k$-modes in total. A Monte Carlo sampling of the system (\ref{Ffk}) was performed in the space of $\nu_k$. The order parameter has been estimated as $\phi^{site}_\nu$ averaged over the Monte Carlo fluctuations of $\nu$.
 The result shown in Figure \ref{natroih} agrees well with the reference data everywhere including 
the critical region.

\begin{figure}[t]
 \center{\includegraphics[width=1\linewidth]{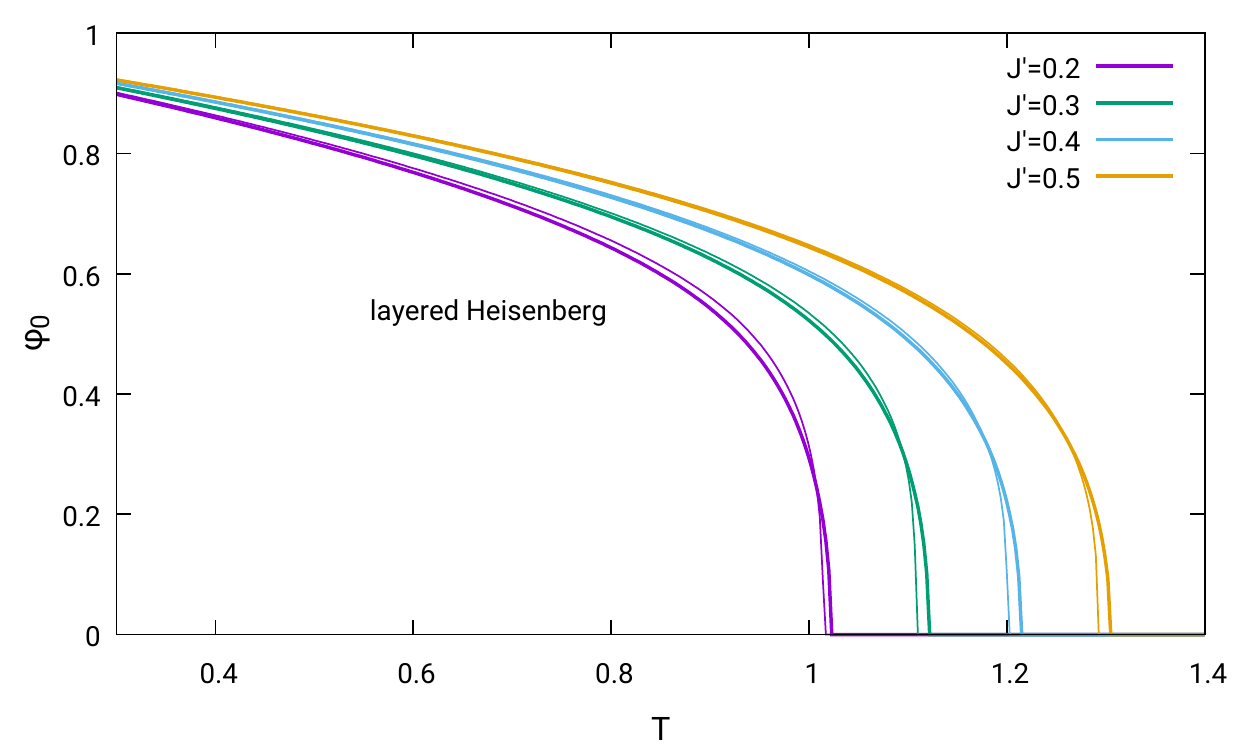}} 
\centering{}\caption{Temperature dependence of the order parameter for the layered Heisenberg model at 
different interlayer coupling. Thick and thin lines correspond to  the cluster methods with $7\times 7$ and $8\times 8$ clusters, respectively. The FLF-2 scheme is used as a cluster solver.}
\label{quarta} 
\end{figure}
 
Finally we provide an example of possible FLF application as a cluster solver. We consider a layered Heisenberg model in which the coupling constant between the layers $J'$ is remarkably less then coupling $J=1$ within the layer. While lowering the temperature, the correlations between the sites belonging to the same layer   grow. Then the inter-layer correlations become important and the phase transition occur. This suggests to introduce planar clusters lying within the layers. We consider periodized 2D clusters placed in a self-consistent field $h=2 J' \phi_0$ reflecting presence of the neighboring layers. Like in any other cluster scheme, 
the results are valid if the scheme shows a convergence with respect to the cluster size. This is achieved for large enough clusters.  We use the FLF-2 scheme with a single fluctuation field (\ref{Ff2}) to obtain the cluster polarization. The Figure \ref{quarta} shows temperature dependence of the order parameter for several values of $J'$, obtained with the $7\times 7$ and $8\times 8$ cluster schemes. The transition temperature is decreased from approximately $1.30 J$ to $1.02 J$ as $J'$ decreases from $0.5 J$ to $0.2 J$. As one can see, the dependence on the cluster size in indeed weak.

\section{Conclusion} 

To summarize, we have considered the thermal equilibrated  finite classical lattices as the simplest systems where the conventional Local Field paradigm is not suitable because of strong collective fluctuations. We have
 presented a Fluctuation Local field approach, which describes 
an interplay of different energy scales, including fluctuations in low energy modes. 
It can be seen as a generalization of the Local Field approach, in which a constant self-consistent field 
of the impurity model is replaced with  the ``slow'' fluctuating field $\nu$. Taking the second-order correction into account allows to use the FLF method as an efficient solver for the cluster schemes with relatively large clusters.

{\it Acknowledgment} The  work was funded by the RSF grant 16-42-01057.

\end{document}